\documentclass[superscriptaddress,
amssymb,amsmath, twocolumn, nobibnotes,aps,pre,showkeys,showpacs]{revtex4-1}
\usepackage{bm}
\usepackage{latexsym}
\usepackage{dcolumn}
\usepackage{amsfonts,amssymb,amsmath}
\usepackage{graphicx,epsfig}
\usepackage{psfrag}
\usepackage{subfigure}
\usepackage{color}
\usepackage{ulem}
\newcommand{\hh}{\widehat{H}}
\newcommand{\vv}{\widehat{V}}
\newcommand{\ff}{\widehat{F}}
\newcommand{\uu}{\widehat{U}}
\newcommand{\uue}{\widehat{\mathcal U}}
\newcommand{\jj}{\widehat{J}}
\newcommand{\fl}{\widehat{\mathcal F}}
\newcommand{\be}{\begin{equation}}
\newcommand{\ee}{\end{equation}}

\begin{document}

\title{Effective time-independent analysis for quantum kicked systems}
\author{Jayendra N. Bandyopadhyay}\email{jnbandyo@gmail.com} \affiliation{Department of Physics, Birla Institute of Technology and Science, Pilani
333031, India.}
\author{Tapomoy Guha Sarkar}\email{tapomoy1@gmail.com} \affiliation {Department of Physics, Birla Institute of Technology and Science, Pilani 333031, India.} 

\begin{abstract}

We present a mapping of potentially chaotic time-dependent quantum
kicked systems to an equivalent approximate effective time-independent
scenario, whereby the system is rendered integrable. The
time-evolution is factorized into an initial kick, followed by an
evolution dictated by a time-independent Hamiltonian and a final
kick. This method is applied to the kicked top model. The effective
time-independent Hamiltonian thus obtained, does not suffer from
spurious divergences encountered if the traditional
Baker-Cambell-Hausdorff treatment is used. The quasienergy spectrum of
the Floquet operator is found to be in excellent agreement with the
energy levels of the effective Hamiltonian for a wide range of system
parameters. The density of states for the effective system exhibits
sharp peak-like features, pointing towards quantum criticality. The
dynamics in the classical limit of the integrable effective
Hamiltonian shows remarkable agreement with the non-integrable map
corresponding to the actual time-dependent system in the non-chaotic
regime. This suggests that the effective Hamiltonian serves as a
substitute for the actual system in the non-chaotic regime at both the
quantum and classical level.

\end{abstract}


\maketitle

\section{Introduction} 

Quantum systems driven periodically in time are
known to undergo remarkable alterations in their long-time dynamical
evolution \cite{shirley, sambe, maricq, rakovic, hanggi, fishman, dalibard}. This is an area of extensive study. External driving is known to create
non-trivial gauge structures \cite{dalibard-RMP, zhai} and topological
effects \cite{zhang, hasan, lindner, grushin, torres1}. Modulated
driving systems have the potential to fabricate new materials and
phases of matter \cite{zhang, hasan, nagosa, gomes, carusotto,
  rechtsman, jacqmin, tarruell, torres2}. In this scheme, the
modulation of the driving system is used to recreate an effective
static Hamiltonian which is thereby investigated for interesting
features of time-independent systems \cite{scharf1, fishman, mintert,
  dalibard}.

Kicked Hamiltonian systems are widely studied as a prototype model for
classical and quantum chaos \cite{haake, stockman}. The traditional
approach to such periodic systems uses the Floquet analysis to obtain
the quasienergies of the systems. Earlier works \cite{shirley, hanggi,
  brandes1} use the formulation based on the Cambell-Baker-Hausdorff
(CBH) expansion to provide an effective time-independent Hamiltonian
\cite{scharf1}. This method has been used to explore non-chaotic
regimes and study critical quasi-energy states \cite{brandes1}.  A
qualitative change in the wave-function of the ground state on
changing one or more parameters of the Hamiltonian of large
systems characterizes a quantum phase transition (QPT)
\cite{sachdev}. However, non-analytic behavior of excited states has
also been studied \cite{brandes1,cejner01,caprio,cejner02}. The
clustering together of a whole branch of states implies that the
average local level spacing $\Delta$ becomes vanishingly small leading
to a divergence of local density of states ($\Delta^{-1}$). This is a
characteristic feature of excited state quantum phase transition
(ESQPT). The corresponding semiclassical Hamiltonian in these cases
indicate dynamical instability.

We claim that the traditional CBH approach to study such time-dependent
problem have intrinsic flaws and an alternative formulation
\cite{fishman, dalibard} is more robust and accurate for the study of
 kicked systems. In this paper, we have applied this alternative
formulation on quantum kicked systems where the time dependence of the
Hamiltonian is in the form of a series of Dirac-delta pulses. We have
studied the specific case of the kicked top model \cite{haake}. We
show that this alternative formulation allows the effective static
Hamiltonian to accurately mimic the exact evolution for a
significantly large range of parameters and the eigenvalues of the
effective Hamiltonian faithfully trace the exact quasienergies without
any spurious divergences (as observed in \cite{brandes1}) that are
pathologies of the CBH based formulation. In the absence of such
singularities, the effective Hamiltonian obtained in our method can be
used to study divergences in the density of states as signature of
ESQPT for a much wider range of
parameter values. We have also shown remarkable match between the
phase-space dynamics of the classical limits of the exact quantum map
\cite{haake} and symplectic evolution governed by the effective static
Hamiltonian in the non-chaotic regime.

\section{Formalism} 

Generic time-dependent problems where $\hh(t) = \hh_0
+ \vv(t)$ is tackled using the Floquet theory when $\vv(t) = \vv(t +
T)$ is time-periodic with period $T$. The Floquet operator $\fl(t)$
corresponds to the time-evolution unitary operator $\uu(t)$ after one
time-period of the driving potential and has eigenvalues of the form
$\exp(- i\, \phi\, T)$ where $\phi$ are refered to as the quasienergies
of the system. In the traditional approach to find an effective
Hamiltonian for kicked systems where 
\be
\vv(t) = \vv \sum\limits_{n = -\infty}^\infty\,\delta(t-nT)
\label{eq:Vt}
\ee
the Floquet operator is factorized as
\be
\fl = \exp(- i \vv) \exp(- i \hh_0 T) = \exp(- i \hh_{\rm eff} T). 
\ee
In the formalism described in \cite{scharf1} and used in
\cite{brandes1}, the effective Hamiltonian $\hh_{\rm eff}$ is
extracted using the CBH expansion and truncated up to a certain order
in $T \sim 1/\omega$. In the absence of any unique prescription for
splitting the Hamiltonian, the Trotter-CBH method is sensitive to the
initial phase of the driving potential. It is evident that any shift
of the initial time $t_i$ will lead to spurious artifact in $\hh_{\rm
  eff}$. This ambiguity has been discussed in the case of square
two-phase modulation of a static Hamiltonian \cite{dalibard}. The
effective Hamiltonian obtained for the kicked top system is also seen
to exhibit unphysical singularities for a class of parameters
\cite{brandes1}. In view of such criticisms of the earlier works
  that use the CBH expansion, to find the effective static
  Hamiltonians, we adopt the alternative formulation
  \cite{fishman,dalibard} to study Floquet systems.

The periodic potential in Eq. \eqref{eq:Vt} may be expanded in a Fourier series as 
\be 
\vv(t) = \vv_0 + \sum\limits_{n=1}^\infty \, \Bigl(\vv_{n} e^{i n \omega t} + \vv_{-n} e^{- i n \omega t}\Bigr).  
\ee 
The method expresses the evolution operator
$\uu(t_i \rightarrow t_f)$ between initial time instant $t_i$ and
final time instant $t_f = t_i + T$, as a sequence of an initial kick
followed by an evolution under a static Hamiltonian and a final
`micro-motion' \cite{dalibard} 
\be 
\uu(t_i \rightarrow t_f) =
\uue^\dagger(t_f) e^{-i \hh_{\rm eff} T}\, \uue(t_i)
\label{eq:floquet}
\ee
where $\uue(t) = e^{i \ff(t)}$ such that $\ff(t) = \ff(t+T)$ with zero average over one time period. In our analysis with
$t_i = 0$ and $t_f = T$, we have $\ff(t_i) = \ff(t_f)$. Thus the
effect of $\uue(t_i) = \uue(t_f)$ is that of similarity transformation
on $\exp(-i\hh_{\rm eff})$. The eigenvalues of $\hh_{\rm eff}$ shall
hence mimic the quasienergies obtained from the Floquet operator. The operators $\hh_{\rm eff}$ and $\ff(t)$ are expanded in a perturbation series in powers of $1/\omega$ as 
\be
\hh_{\rm eff} = \sum\limits_{n = 0}^\infty\, \frac{1}{\omega^n} \hh_{\rm eff}^{(n)},~~~~\ff(t) = \sum\limits_{n = 1}^\infty\, \frac{1}{\omega^n} \ff^{(n)}.
\ee 
Comparing this with Eq. \eqref{eq:floquet}, the perturbation series can be obtained up to any desired accuracy. At each order of perturbation, the averaged time-independent component is retained in $\hh_{\rm eff}$ and all time-dependence is pushed into the operator $\ff(t)$. This yields up to $\mathcal{O}(1/\omega^2)$ for the Hamiltonian $\hh(t)$ \cite{dalibard}:

\begin{widetext}
\be
\begin{split}
\hh_{\rm eff} &= \hh_0 + \vv_0 + \frac{1}{\omega}
\sum\limits_{n=1}^\infty \frac{1}{n}\bigl[\vv_n, \vv_{-n}\bigr] +
\frac{1}{2\omega^2}\sum\limits_{n=1}^\infty \frac{1}{n^2} \Bigl(
\bigl[\bigl[ \vv_n, \hh_0\bigr], \vv_{-n}\bigr] + {\rm h.c.}\Bigr)\\
& + \frac{1}{3\omega^2} \sum\limits_{n, m = 1}^\infty \frac{1}{nm}
\Bigl(\bigl[\vv_n, \bigl[\vv_m, \vv_{-n-m}\bigr]\bigr]\Bigr.  \Bigl.-
2 \bigl[\vv_n, \bigl[\vv_{-m}, \vv_{m-n}\bigr]\bigr] + {\rm
  h.c.}\Bigr) \\ \ff(t) &=
\frac{1}{i\omega}\displaystyle\sum_{n=1}^{\infty}\frac{1}{n}\Bigl(\vv_ne^{in\omega
  t}-\vv_{-n}e^{-in\omega t}\Bigr) +
\frac{1}{i\omega^2}\displaystyle\sum_{n=1}^{\infty}\frac{1}{n^2}\Bigl(\bigl[\vv_n,\hh_0+\vv_0
  \bigr]e^{in\omega t} - {\rm h.c.}\Bigr)\\ &+ \frac{1}{2i\omega^2}
\sum\limits_{n, m = 1}^\infty \frac{1}{n (n+m)}
\Bigl(\bigl[\vv_n,\vv_m] e^{i(n+m)\omega t} - {\rm h.c.}\Bigr)+
\frac{1}{2i\omega^2}\sum\limits_{n\ne m = 1}^\infty \frac{1}{n (n-m)}
\Bigl(\bigl[\vv_n,\vv_{-m}\bigr] e^{i(n-m)\omega t} - {\rm
  h.c.}\Bigr).
\end{split}
\label{eq:hamnf}
\ee 
\end{widetext}
For general quantum kicked systems with Dirac-delta forcing, $\vv_n = \vv/T$ for all $n = 0, \pm 1, \pm 2, \dots, \pm \infty$. Therefore, the above expression can be simplified as
\begin{widetext}
\be
\begin{split}
&\hh_{\rm eff} = \hh_0 + \frac{\vv}{T} + \frac{1}{\omega^2 T^2} \bigl[\bigl[\vv, H_0], \vv\bigr] \left(\sum\limits_{n=1}^\infty \frac{1}{n^2}\right)
= \hh_0 + \frac{\vv}{T} + \frac{1}{24} \bigl[\bigl[\vv, H_0], \vv\bigr]\\
\ff(t) = \frac{2 \vv}{\omega T} & \sum\limits_{n=1}^\infty \frac{\sin(n\omega t)}{n} + \frac{2}{i \omega^2 T} \bigl[\vv, \hh_0\bigr] \sum\limits_{n =1}^\infty \frac{\cos(n\omega t)}{n^2} = \frac{\vv}{\pi}\, \sum\limits_{n=1}^\infty \frac{\sin(n\omega t)}{n} - i\, \frac{T}{2 \pi^2}\, \bigl[\vv, \hh_0 \bigr] \sum\limits_{n =1}^\infty \frac{\cos(n\omega t)}{n^2}.
\end{split}
\ee 
\end{widetext}
It is evident that in the partitioning of the time-evolution operator as
in Eq. \eqref{eq:floquet} the system is assumed to undergo an initial
kick $\exp[i \ff(t_i)]$ which is sensitive to the launching time
$t_i$. This, therefore has a long term bearing on the dynamical
evolution, though the evolution after the initial kick is essentially
dictated by the static effective Hamiltonian. 

\section{Kicked top model}

The kicked top model is representative of a host of such kicked
systems and manifests chaotic dynamics
\cite{haake-paper1,haake-paper2,haake-paper3}. This model is also
known to have closed bearings with condensed matter systems like
metal-topological-insulator \cite{altland, beenakker} and has
also been studied in the context of quantum critical transition
\cite{brandes2,brandes1}. The study of the kicked top is also
motivated by recent experiments \cite{saumya1,saumya2}.

We consider the Hamiltonian for the kicked top
\be
\hh(t) = \frac{\alpha}{2 j T} \jj_z^2 + \beta \jj_x \sum\limits_{n = -\infty}^\infty\,\delta(t-nT),
\label{eq:ham}
\ee where $\jj_i$ denotes the components of the angular momentum, $j
(j+1)$ is the eigenvalue of $\jj^2$. The operators $\jj_z$ and $\jj_x$ are
the $z-$ and $x-$components of the angular momentum operator,
respectively. Three components of the angular momentum operator
satisfy standard commutation relation $[\jj_i,\jj_j] = i \epsilon_{ijk}
\jj_k$. The Floquet operator for the above Hamiltonian is given by \be
\fl = \exp(-i\beta \jj_x) \exp\left(-i\frac{\alpha \jj_z^2}{2jT}\right).
\ee The first factor in $\fl$ describes a rotation operator around the
$x$-axis by the angle $\beta$. The second factor corresponds to a
nonuniform rotation or torsion around the $z$-axis. This term consists of a 
 rotation angle  which is itself proportional to the angular momentum
component $J_z$. Therefore, the parameter $\alpha$ measures the
torsional strength. 

The perturbation expansion is tracked using the tracking parameter $1/\omega$.
Evaluating the commutators in Eq. \eqref{eq:hamnf}, we have the expressions for the truncated $\hh_{\rm eff}$ and $\ff$ up to $\mathcal{O}(1/\omega^2)$ given by
\be
\begin{split}
\hh_{\rm eff} & = \frac{\alpha}{2j} \jj_z^2 + \beta \jj_x - \frac{\alpha \beta^2}{24 j} (\jj_z^2 - \jj_y^2)\\
\ff(t_i) & = \ff(t_f)  = -\frac{\alpha \beta}{24j} \bigl(\jj_y \jj_z + \jj_z \jj_y\bigr).
\label{ham_eff} 
\end{split}
\ee

\begin{figure}[ht]
\includegraphics[height=6.0cm, width=8.5cm]{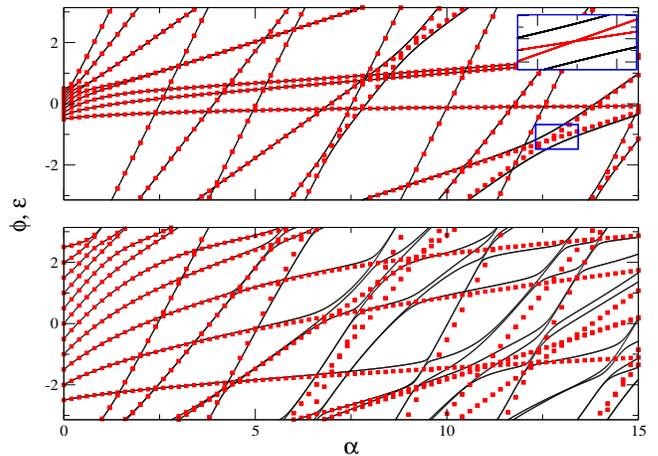}
\caption{(Color online) Quasienergy spectrum (solid line) of the Floquet operator $\fl$ and the energy eigenvalues (solid square) of the effective Hamiltonian $\hh_{\rm eff}$ are compared as a function of $\alpha$. Upper window shows the result for $\beta = 0.1$, and the lower window shows the same for $\beta = 0.5$. Whereas the integrable $\hh_{\rm eff}$ shows level crossing, the quasienergy spectrum indicates level repulsion (see inset).}
\label{fig:fig1}
\end{figure}

\noindent Here we have truncated the perturbation series at the second order  and subsequently used the units of time such that $\omega = 2\pi$. 

We note that $\hh_{\rm eff}$ does not show any singularities for any
values of the parameters $\alpha$ and $\beta$. This feature differs
crucially from the results obtained earlier \cite{brandes1} where the
effective quasienergies showed divergence when $\alpha (2m+1) = 4 j l
\pi$ where $l \in \mathbb{Z}$ and $m$ denotes the eigenvalues of
$\jj_z$. We claim that these singularities are an artifact of the
naive use of the CBH formula.

\subsection{Quasienergy Spectrum} 

Figure \ref{fig:fig1} shows the numerically obtained eigenvalues of the Floquet operator $\fl$ and the
effective Hamiltonian $\hh_{\rm eff}$ for $\beta=0.1$. We note that,
the former is non-integrable whereas the later is static, approximate,
and integrable. The figure shows the eigenvalues $\epsilon$ of
$\hh_{\rm eff}$ mapped into the first Brillouin zone. A remarkable
match of $\epsilon$ with the quasienergies $\phi$ is obtained for broad
range of values of $\alpha$. The agreement of the approximate
eigenvalues with the exact quasienergies at a very high level of
precision is noticed even in the domain of the parameters $\alpha$ for
which the system approaches chaotic regime. This is different from the
results obtained in an earlier work \cite{brandes1} where such
agreement is noticed for a very small range of values of $\alpha$.

 The alternative formulation yields new physical insight regarding the
 use of approximate methods to deal with time-dependent systems. The
 splitting of the time-evolution operator into an initial kick, a
 final kick, and an intermediate time-evolution dictated by a
 stationary Hamiltonian, is expected to give better results if the
 original time-dependent Hamiltonian itself is comprised of periodic
 kick pulses as in the present case of the kicked top. The idea has
 been to shift the effects of these kicks to the initial and final
 moments of time-evolution by means of a unitary transformation. The
 success of this formulation implies that the method should be used
 for the generic class of kicked systems as against the CBH based
 method.

 The fundamental departure in the approximate analysis adopted here, occurs
around regions where the exact quasienergy spectrum shows level
repulsion which is characteristic of non-integrable systems. In the
eigenspectrum of $\hh_{\rm eff}$ which is an integrable system there
is manifestation of degeneracy. The actual quasienergy spectrum avoids
such crossings \cite{scharf2}. The situation gets worse for higher
values of $\beta$ where more of such spurious crossings appear.

\begin{figure}[hb]
\includegraphics[height=7.0cm,width=8.5cm]{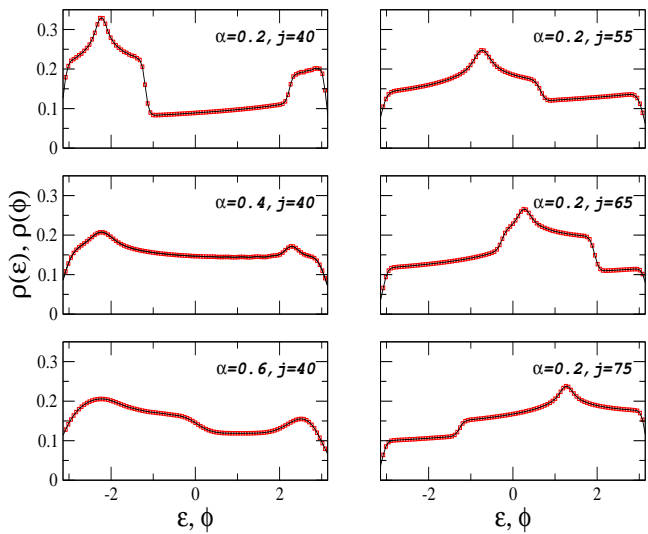}
\caption{(Color online) Density of states of the exact quasienergy
  spectrum of $\fl$ and of the energy spectrum of $\hh_{\rm eff}$ are
  compared. Here the parameter $\beta = 0.1$.}
\label{fig:fig2}
\end{figure}

 To investigate the nature of critical quasienergy states for a much
wider range of parameters in the Hamiltonian we compare the density of states (DOS) for the quasienergy spectrum and that of the eigenspectrum of $\hh_{\rm eff}$.
The parameter values for which the energy spectrum indicates a tendency of clustering is also expected to manifest as divergence in the density of states 
\be
\rho(\mathcal{E}) = \frac{1}{2j+1} \sum_r \delta(\mathcal{E}-\mathcal{E}_r)
\label{eq:dos01}
\ee
where $\mathcal{E} = \phi$ and $\mathcal{E} = \epsilon$ for the quasienergy spectrum and the eigenvalues of
$\hh_{\rm eff}$, respectively. This can alternatively be written as 
\be 
\rho(\mathcal{E}) = \frac{1}{2\pi} + \frac{1}{\pi(2j+1)} {\rm Re} \left\{\sum_{n=1}^\infty \xi_n e^{i n \mathcal{E}}\right\}
\label{eq:dos02} 
\ee 
where $\xi_n = \sum_r \exp(- i n \mathcal{E}_r)$. We note that Eq. \eqref{eq:dos02} uses the Fourier representation of Dirac-deltas in Eq. \eqref{eq:dos01}. 

The DOS given in \eqref{eq:dos01} or \eqref{eq:dos02}, for a discrete
spectrum would yield a series of jagged vertical lines corresponding
to the Dirac-deltas. Binning these discrete distribution would yield a
histogram. We have numerically obtained the smoothed DOS by
considering the Gaussian representation of the Dirac-deltas and
obtaining a continuous curve by summing these Gaussian functions
centered at different values of $\mathcal{E}_r$. The widths of the
normalized Gaussians are kept at $\sim 10\%$ of the mean spacing to
ensure smooth and continuous DOS. The same numerics are used to obtain
both the DOS for exact quasienergy spectrum and the eigenvalues of
$\hh_{\rm eff}$. The parameter of the Hamiltonian for which there is a
clustering tendency of the levels shall also reflects as critical
transition point where the DOS is likely to diverge.

Figure \ref{fig:fig2} compares the DOS for the exact quasienergy
spectrum and the energy spectrum of $\hh_{\rm eff}$. The strong
resemblance of these spectra (see Fig. \ref{fig:fig1}) indicates that
their DOS should also match. We find that the two are indeed found to
match to a high degree of accuracy for different values of the
parameter $\alpha$, and also for different spin. The sharp
peak-feature that appears in the DOS for $\alpha=0.2$ for spin $j=40$
indicates quantum criticality as suggested in an earlier work
\cite{brandes1}.  At the critical quasienergy, a clustering of states
occur leading to the tendency of DOS to diverge.

 The feature however is sensitive to $\alpha$ and flattens out at
 larger values of $\alpha$. The position of the peak is seen to shift
 as $j$ increases. This is quite expected since the DOS is closely
 related to the degeneracies of the energy levels which in turn
 depends on $j$. The claim for the indication of quantum criticality
 as reflected in logarithmic divergence of the DOS is noted in an
 earlier work \cite{brandes1}. This result is vindicated by our
 present analysis which uses an effective Hamiltonian intrinsically
 differing from the one used earlier but however, does not suffer from
 mismatches in the spectrum at larger values of $\alpha$.

\subsection{The classical limit}

Using the Heisenberg equation of motion, we find the following quantum dynamical map for the angular momentum operator as
\be
\widehat{\mathbf J}_{n+1} = \fl^\dagger\, \widehat{\mathbf J}_n\,\fl.
\label{eq:quant-dyn}
\ee The classical limit of this map can be achieved by first rescaling
the operator $\widehat{\mathbf J}$ as $\mathbf{X} = \widehat{\mathbf J}/j$, i.e., $\{X, Y, Z\} = \{\jj_x, \jj_y, \jj_z\}/j$, where the commutators of the different components of the rescaled angular momentum operators take the form
$[X, Y] = i Z/j$, and so on. This shows that, in $j \rightarrow
\infty$ limit, components of this rescaled angular momentum operator
will commute and become classical $c$-number variables, and we get the
classical limit of the quantum dynamical map as \be
\begin{split}
X_{n+1} &= \widetilde{X} \cos \alpha \widetilde{Z} - \widetilde{Y} \sin \alpha \widetilde{Z}\\
Y_{n+1} &= \widetilde{X} \sin \alpha \widetilde{Z} + \widetilde{Y} \cos \alpha \widetilde{Z}\\
Z_{n+1} &= \widetilde{Z},~~~~~ {\rm where}\\
\widetilde{X} &= X\\
\widetilde{Y} &= Y \cos \beta - Z \sin \beta\\
\widetilde{Z} &= Y \sin \beta + Z \cos \beta. 
\end{split}
\label{eq:class_map}
\ee The above map satisfies the condition $X^2 + Y^2 + Z^2 = 1$. This
suggests that the classical phase-space dynamics of the kicked top lies on
the surface of a unit sphere, and each point on that surface is
represented by two canonically conjugate dynamical variables $Z = \cos
\theta$ and $\psi = \tan^{-1}(Y/X)$.

This classical map is to be compared with the dynamical solution in the phase-space corresponding to the classical limit of the effective Hamiltonian $\hh_{\rm eff}$ in Eq. \eqref{ham_eff}. We use the following prescription to find this classical limit designated by $H_{\rm cl}$ with 
\be
H_{\rm cl} = \lim_{j \rightarrow \infty} \frac{\langle\gamma|\hh_{\rm eff}|\gamma\rangle}{j}
\ee
where $|\gamma\rangle$ is the spin coherent state \cite{haake}. From Eq. \eqref{ham_eff}, this yields 
\begin{widetext} 
\be
H_{\rm cl} = \frac{\alpha Z^2}{2} + \beta \sqrt{1-Z^2} \cos\psi + \frac{\alpha \beta^2}{24} \sin^2\psi - \frac{\alpha \beta^2}{24} (1+\sin^2\psi) Z^2
\ee
\end{widetext}
This classical Hamiltonian represents the effective integrable model corresponding to the original Floquet system. The form of this Hamiltonian shows that it is non-separable. The terms contain the canonical variables $(Z, \psi)$ in a manner which do not allow a separation of $H_{\rm cl}$ into a purely position/momentum dependent components. Hamilton's canonical equations are given as follows: 
\be
\begin{split}
\dot{Z} &= -\frac{\partial H_{\rm cl}}{\partial \psi} = \beta \sqrt{1-Z^2} \sin \psi - \frac{\alpha\beta^2}{24} (1-Z^2)\sin 2\psi\\
\dot{\psi} &= \frac{\partial H_{\rm cl}}{\partial Z} = \alpha Z - \frac{\beta Z}{\sqrt{1-Z^2}} \cos\psi - \frac{\alpha\beta^2}{12} Z (1+\sin^2 \psi).
\end{split}
\label{eq:eqn_motion}
\ee
The symplectic evolution for this dynamical system mimics the map in Eq. \eqref{eq:class_map} despite the fundamental difference between the two situations. The former represents conservative evolution satisfying Liouville's theorem without any signature of chaos for any values of the parameters $\alpha$ and $\beta$. The latter, on the contrary, is a well studied candidate of quantum and classical chaos for large $\alpha, \beta \gg 1$. 

The equation \eqref{eq:eqn_motion} can be also recast in terms of the variables $\{X, Y, Z\}$ as 
\be
\begin{split}
\dot{X} &= - \left(\alpha-\frac{\alpha\beta^2}{6}\right) Y Z\\
\dot{Y} &= \left(\alpha-\frac{\alpha\beta^2}{12}\right) X Z - \beta Z\\
\dot{Z} &= \beta Y - \frac{\alpha\beta^2}{12} X Y.
\end{split}
\label{eq:eqn_Poiss}
\ee
We note that this equation can be obtained as the large $j$ limit of the quantum Hamiltonian 
\be
\hh/j = (\alpha/2)\widehat{Z}^2 + \beta \widehat{X} + (\alpha\beta^2/24)(\widehat{Z}^2 - \widehat{Y}^2)
\ee
using the Heisenberg's equation of motion for the operators $\{\widehat{X}, \widehat{Y}, \widehat{Z}\}$, and subsequently replacing the quantum commutators by classical Poisson brackets. 

\begin{figure}[h]
\includegraphics[height=6.5cm,width=8cm]{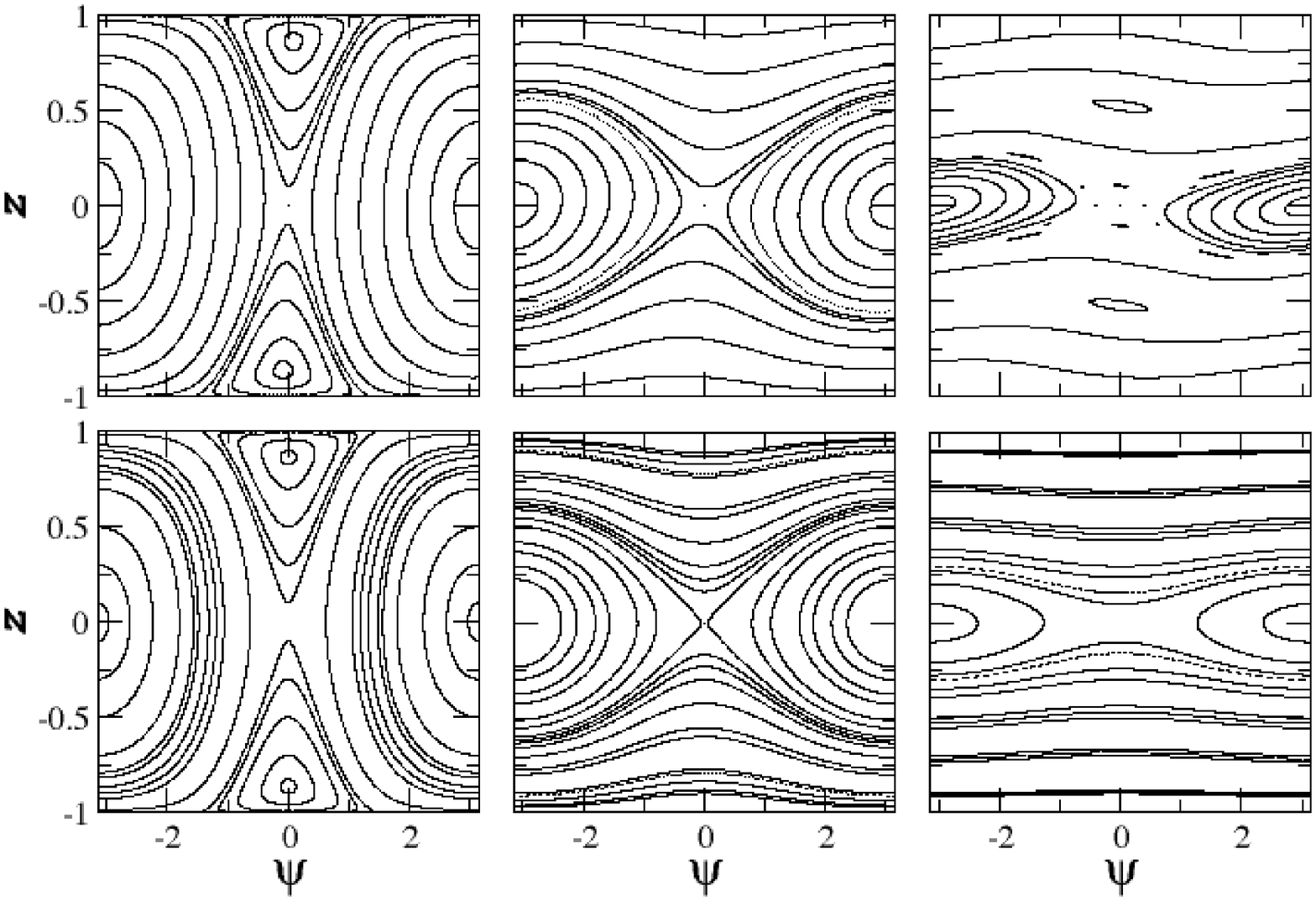}
\caption{(Color online) The upper panel shows the phase-space corresponding to the non-integrable evolution governed by the map given in Eq. \eqref{eq:class_map} for parameter $\beta = 0.1$. The lower panel shows the corresponding phase for the effective time-independent integrable Hamiltonian in Eq. \eqref{eq:eqn_Poiss}. The parameter $\alpha$ is increasing from left to right as: $\alpha = 0.2$ (left column), $1.0$ (middle column), and $6.0$ (right column). From left to right, the upper panel shows the system evolving towards chaotic regime. There is no such indication for the integrable system depicted in the lower panel.}  
\label{fig:fig3}
\end{figure}

Figure \ref{fig:fig3} shows the phase space for the dynamical systems
represented by Eq. \eqref{eq:eqn_motion} and the classical map obtained
for the actual time-dependent Floquet system described in
Eq. \eqref{eq:class_map}. In the predominantly non-chaotic regular
regime for small $\alpha, \beta$ values, the phase space trajectories
agree to a very high degree and the effective time-independent
classical Hamiltonian mimics the actual system for most purposes
pertaining to dynamics. The departure begins to show up as the actual
time-dependent system approaches chaotic regime, and the trajectories,
despite reflecting the same generic form, do not quite follow the same
path. As the actual non-integrable system approaches the chaotic
regime the phase-space shows the formation of islands which eventually
break into further substructures. Expectedly, this feature is
completely missing in the dynamics of the effective integrable
Hamiltonian, where the phase-space always remains regular and thereby
cannot represent the actual time-dependent system for large parameter
values. We note that there are studies indicating the disappearance of
island like features in certain time-independent nonintegrable systems
when they are approximated by integrable models \cite{ketzmerick}.

\section{conclusion}

We conclude by noting that by introducing an unitary transformation
that pushes the time-dependence of a time-periodic system to the
initial time and final time instants in the form of kicks leads to an
effective time-independent Hamiltonian that governs evolution for the
bulk of the time. The energy spectrum of this effective Hamiltonian
matches with the exact quasienergies of the actual Floquet operator in the
near integrable regime with deviations due to avoided crossings.
We also find that the classical limit mimics the phase-space dynamics of the
actual time-dependent non-integrable Hamiltonian in the non-chaotic
regime.

The CBH based method has been used extensively
in the scientific literature  without any
concerns about its validity in a given context. The present work
illustrates the fact that there are situations where the
indiscriminate use may lead to incorrect conclusions.
We also establish that the formulation adopted by us is probably a more
appropriate approximate analysis of classical and quantum systems 
involving short duration pulse driving.

\begin{acknowledgments}
Authors would like to thank Prof. A. Lakshminarayan of IIT-Madras, India for valuable comments. A special thank to Mr. Tridev Mishra for useful discussions.
\end{acknowledgments}


\begin{thebibliography}{99}

\bibitem{shirley} J. H. Shirley, Phys. Rev. {\bf 138}, B979 (1965).
\bibitem{sambe} H. Sambe, Phys. Rev. A {\bf 7}, 2203 (1973).
\bibitem{maricq} M. M. Maricq, Phys. Rev. B {\bf 25}, 6622 (1982).
\bibitem{rakovic} T. P. Grozdanov and M. J. Rakovic, Phys. Rev. A {\bf 38}, 1739 (1988).
\bibitem{hanggi} M. Grifoni and P. H\"anggi, Phys. Rep. {\bf 304}, 229 (1998).
\bibitem{fishman} S. Rahav, I. Gilary, and S. Fishman, Phys. Rev. A {\bf 68}, 013820 (2003).
\bibitem{dalibard} N. Goldman and J. Dalibard, arXiv:1404.4373v2.
\bibitem{dalibard-RMP} J. Dalibard et. al., Rev. Mod. Phys. {\bf 83}, 1523 (2011).
\bibitem{zhai} Int. J. Mod. Phys. B {\bf 26}, 1230001 (2012). 
\bibitem{zhang} X. L. Qi and S. C. Zhang, Rev. Mod. Phys. {\bf 83}, 1057 (2010).
\bibitem{hasan} M. Hasan and C. Kane, Rev. Mod. Phys. {\bf 82}, 3045 (2010).
\bibitem{lindner} N. H. Lindner, G. Refael, and V. Galitski, Nat. Phys. {\bf 7}, 490 (2011).
\bibitem{grushin} A. G. Grushin, A. G\'omez-Le\'on, and T. Neupert, Phys. Rev. Lett. {\bf 112}, 156801 (2014).
\bibitem{torres1} P. M. Perez-Piskunow et. al., Phys. Rev. B {\bf 89}, 121401(R) (2014).
\bibitem{nagosa} P. A. Lee, N. Nagosa, and X. G. Wen, Rev. Mod. Phys. {\bf 78}, 1057 (2006).
\bibitem{gomes} K. K. Gomes et. al., Nature {\bf 8}, 483 (2012).
\bibitem{tarruell} L. Tarruell et. al., Nature {\bf 483}, 302 (2012).
\bibitem{torres2} E. S. Morell and L. E. F. Foa Torres, Phys. Rev. B {\bf 86}, 125449 (2012).
\bibitem{carusotto} I. Carusotto and C. Ciuti, Rev. Mod. Phys. {\bf 85}, 299 (2013).
\bibitem{rechtsman} M. C. Rechtsman et. al., Nature {\bf 496}, 196 (2013). 
\bibitem{jacqmin} T. Jacqmin et. al., Phys. Rev. Lett. {\bf 112}, 116402 (2014)
\bibitem{mintert} A. Verdeney, A. Mielke, and F. Mintert, Phys. Rev. Lett. {\bf 111}, 175301 (2013).
\bibitem{scharf1} R. Scharf, J. Phys. A {\bf 21}, 2007 (1988) 
\bibitem{haake} F. Haake, {\it Quantum Signature of Chaos} (Springer, Berlin, 2009), 3rd Ed.
\bibitem{stockman} H. -J. St\"ockman, {\it Quantum Chaos: An Introduction} (Camb. Univ. Press, London, 1999)
\bibitem{brandes1} V. M. Bastidas et. al., Phys. Rev. Lett. {\bf 112}, 140408 (2014). 
\bibitem{sachdev} S. Sachdev, {\it Quantum Phase Transitions} (Canbridge University Press, Cambridge, England, 1999).
\bibitem{cejner01} P. Cejnar, M. Macek, S. Heinze, J. Jolie, and J. Dobes, J.
Phys. A {\bf 39}, L515 (2006).
\bibitem{caprio} M. A. Caprio, P. Cejnar, and F. Iachello, Ann. Phys. (N.Y.)
{\bf 323}, 1106 (2008).
\bibitem{cejner02} P. Cejnar and P. Str‡nsk\'y, Phys. Rev. E 78, 031130 (2008).
\bibitem{scharf2} R. Scharf, J. Phys. A {\bf 21}, 4133 (1988).
\bibitem{haake-paper1} F. Haake, M. Kus, and R. Scharf, Z. Phys. B {\bf 65}, 381 (1987).
\bibitem{haake-paper2} M. Kus, F. Haake, and B. Eckhardt, Z. Phys. B {\bf 92}, 221 (1993).
\bibitem{haake-paper3} P. Gerwinski et. al., Phys. Rev. Lett. {\bf 74}, 1562 (1995). 
\bibitem{altland} C. Tian, A. Altland, and M. Garst, Phys. Rev. Lett. {\bf 107}, 074101 (2011). 
\bibitem{beenakker} E. P. L. van Nieuwenburg et. al., Phys. Rev. B {\bf 85}, 165131 (2012)
\bibitem{brandes2} T. Brandes, Phys. Rev. E {\bf 88}, 032133 (2013).
\bibitem{saumya1} S. Chaudhury et. al., Phys. Rev. Lett. {\bf 99}, 163002 (2007).
\bibitem{saumya2} S. Chaudhury et. al., Nature {\bf 461}, 768 (2009).
\bibitem{ketzmerick} C. L\"obner et. al., Phys. Rev. E {\bf 88}, 062901 (2013) 

\end{thebibliography}
\end{document}